# Insulating band gaps both below and above the Néel temperature in d-electron LaTiO$_3$, LaVO$_3$, SrMnO$_3$, and LaMnO$_3$ perovskites as a symmetry-breaking event

Oleksandr I. Malyi, Xin-Gang Zhao, and Alex Zunger

Renewable and Sustainable Energy Institute, University of Colorado, Boulder, Colorado 80309, USA

Metal d-electron oxides having an odd number of electrons per cell should exhibit band degeneracy at the Fermi energy, making them, in band theory, formally metallic. In many cases, however, these are "false metals", as evidenced by the observation that many ABO$_3$ oxide perovskites with a magnetic 3d B-atom are observed to be insulators both below and above the Néel temperature. These inconsistencies between experimental observation and expectation have historically been resolved by invoking degeneracy-breaking physics, based largely on—pure *electron effects,* such as strong interelectronic correlation for d-electron compounds (the Mott mechanism). Such explanations generally consider microscopic lattice or magnetic degrees of freedom (m-DOF) as largely passive spectators, not a cause of the formally metallic being an insulator. Yet, it has long been known that ABO$_3$ perovskites can manifest an arrangement of m-DOFs in the form of octahedral tilting, bond dimerization, Jahn-Teller distortions, and ordering of local magnetic moments. It appears reasonable that such structural and magnetic local degrees of freedom need to be allowed to compete with purely electronic strong correlation. To answer this question, we explored a range of d-electron oxide perovskites exemplified by the archetypes LaTiO$_3$, LaVO$_3$, SrMnO$_3$, and LaMnO$_3$ with 1, 2, 3, or 4 d-electrons, respectively. Using a mean-field-like electronic structure method (here, density functional theory), we find that a combination of magnetic symmetry breaking with structural distortions can account for insulating band gaps in this series while at the same time correctly predicting for the control case an intrinsic paramagnetic metal in SrVO$_3$ as symmetry-breaking is insufficiently strong to remove the degeneracy. This indicates that calculating quantitatively local magnetic and positional symmetry-breaking motifs in unit cells that avoid averaging at the outset over the low symmetry motifs can provide consistent trends in a "Mott transition without Mott U".

# I. Introduction

Compounds that exhibit the same orbital characteristics for both filled and unfilled band edge states - for instance, when both are d-like and the number of electrons is odd - are predicted to show band degeneracy (metallic properties) at the Fermi energy level [1-4]. The conceptual usefulness of such reference systems stems from the fact that the experimentally observed systems supposed to be close to such paradigm reference have turned out in reality to be insulators. This was the case not just for the magnetically long-rang ordered ground state phases (i.e., antiferromagnetic phases below the Néel temperature), but, surprisingly, also above the Néel temperature in the paramagnetic (PM) phases, which lacks magnetic long-range order. Such insulating phases were seen in a range of d-electron oxide perovskites $ABO_3$ exemplified by the archetypes $LaTiO_3$, $LaVO_3$, $SrMnO_3$, and $LaMnO_3$ with 1, 2, 3, or 4 d-electrons, respectively, contributed by the B ion, rather than the theoretically imagined "false metal" state [5]. This conflict has set the stage for the historical pursuit of the crucial "missing ingredient" in our understanding of the false metal being instead a true insulator both below and above the Néel temperature in such compounds. What made this conundrum even more intriguing was the fact that $SrVO_3$—the single d- electron isovalent analog to insulating $LaTiO_3$ – has been shown to persistently be true metal at all temperatures[6,7], and that $YNiO_3$ is a true insulator in the low T phase, becoming a true metal in the high T phase[8].

The common textbook approach for resolving such inconsistencies between experimental observation and model expectations has followed the route of "Electron Phases of Matter", focusing largely on the role of electronic degrees of freedom. The degeneracy breaking physics considered in such Mott insulator with (d,d*) band edges, for example, has focused largely on strong electron correlation, encoded, for example, by the on-site Coulomb repulsion U. When sufficiently strong, it could localize electrons, creating a (Mott) insulator instead of a false metal. The microscopic *lattice* degrees of freedoms (m-DOFs)—such as octahedral tilting, bond dimerization, Jahn-Teller distortions, octahedra disproportionation and ordering of local magnetic moments – were largely considered as spectator degrees of freedom, rather than the cause of the false metal reference system being actually an insulator.

This "strong correlation" viewpoint of gap formation is generally inaccessible to mean-field-like band structure methods, such as Density Functional Theory (DFT). This led to the view that a strong correlation is a necessary factor needed for the formation of real insulators from false metals. This position of the strongly correlated literature has been voiced for a long time, including very recently for the $d^1$, $d^2$, $d^3$, and $d^4$ compounds $LaTiO_3$[9,10], $LaVO_3$[11], $SrMnO_3$[12], and $LaMnO_3$[13], respectively. For instance, Gull et al. indicated lately [12] regarding $SrVO_3$ and $SrMnO_3$ that "standard electronic structure methods such as DFT and GW are unable to reproduce it due to the missing correlations in their partially filled transition metal shells. The quasiparticle bandwidth in $SrVO_3$ is too wide, and $SrMnO_3$ is metallic rather than insulating". A similar impression was echoed again recently by Pavarini, pointing[13] out that $LaMnO_3$ was found to be a metal in the Kohn–Sham version of DFT band theory, but it is an insulator, in reality, adding, "there are entire classes of materials for which this practice fails qualitatively due to strong local electron-electron repulsion effects".

The question pondered here regards what is the minimal physics needed to describe the basic trends in the phases noted above. The low temperature long-range ordered "ground state" phases of $ABO_3$ are normally described by crystallographic unit cells that can geometrically accommodate non-trivial configurations of m-DOFs. The ensuing band structure calculations then reflect the electronic consequences of these m-DOFs, often representing local symmetry breaking motifs. The reason that m-DOFs were often dismissed as a potential explanation of the insulating state *above* the Néel temperature was the tradition of describing such phases in band theory as the highest symmetry structure, after the local motifs have, in fact, been "averaged out". Phases that lack long-range order of m-DOFs such as idealized paramagnetic, paraelectric or paraelastic were indeed often simplistically described by high symmetry space groups. For example, the nominal cubic Pm-3m space group common in many perovskite structures contains but a single $ABO_3$ repeat unit, thus unable to geometrically describe symmetry breaking. Band structure calculations of such high symmetry structures cannot examine degeneracy removal by m-DOFs. However, the absence of long-range order of local motifs in such phases does not mean that short range order is absent. Avoiding the use of average, high symmetry unit cell for para phases or cubic phases would allow instead the possibility of a polymorphous network where a distribution of local magnetic of structural motifs would exist, should it lower the total energy. It appears reasonable that such structural and magnetic local degrees of freedom need to be allowed to compete with purely electronic strong correlation in examining if the correct insulating vs metallic phases would emerge as a result of local symmetry breaking. It is worth considering if the intrinsic structural and magnetic symmetry breaking (present even before temperature sets in) might be related to the formation of insulating band gaps both below and above the magnetic transition, as observed in $LaTiO_3$, $LaVO_3$, $SrMnO_3$, and $LaMnO_3$. Here, we explore the outcome of such a back-and-forth "ping-pong" sequence where the electronic structure responds to (non averaged) m-DOFs, and in turn, the modified electronic landscape can reshape these m-DOF properties. To pinpoint the key factor determining whether a material is an insulator or metal in a given phase, various potential symmetry-breaking pathways[14] are considered. This requires avoiding the imposed high symmetry "virtual lattice averages" over the local motifs both below and above the Neel temperature. Instead, each macroscopic phase is described by a "supercell" of its minimal cells, thereby allowing the possibility of polymorphous distribution of *local* motifs within the globally fixed crystallographic structure.

The main conclusions noted are:

(i) The basic metal vs. insulator observed phenomenology described above is consistently predicted by mean field like band theory if the local positional and magnetic symmetry breaking are examined. This symmetry breaking predicts (a) insulating band gaps both below and above the magnetic transition, as observed in $LaTiO_3$, $LaVO_3$, $SrMnO_3$, and $LaMnO_3$. Such insulating phases can occur in both magnetically long-range ordered (LRO) phases (below the Néel temperature)[15] and in PM phases that lack magnetic LRO[5,16-20]. (b) Additionally, the same theoretical approach applies to the "control case" of perovskite compounds which appear electronically and structurally indistinguishable from the $d^1$ insulator subgroup, but in fact are observed to be *persistent metals*. This was found to be the case whether the relevant d-orbital is a compact 3d (as $d^1$ SrVO$_3$)[6,7] or a delocalized 4d states (as in SrNbO$_3$ and BaNbO$_3$)[21,22]. (c) Also, the recent application of the same method to YNiO$_3$[23] (with temperature introduced via DFT Molecular Dynamics) explains the prototype class of insulating below the Neel temperature and metallic above it.

(ii) The distinction between this work and the earlier works on d-electron oxide perovskites is that two local symmetry-breaking mechanisms – positional and magnetic – are allowed here to operate together to achieve an insulating state out of a reference zero gap metal. Comparable absolute increase in the magnitude of band gaps were noted[24-26] (in cubic, non-magnetic and non-d electron *halide* perovskites such as CsPbX$_3$, where X is a halogen). However, the initial band gap in the halide case is already non zero.

(iii) The common belief about DFT that it cannot handle the Mott transition – a claim that has been widely exploited to introduce the dynamic mean-field theory (DMFT) [12,13] – cannot really be use for these classic 3d oxides as a motivation for this claim.

(iv) The physics of correcting the false metal expectation does not have to rely on the on-site Coulomb repulsion Mott-Hubbard view of strong correlation. Indeed, symmetry breaking allows such a Mott transition without Mott U. Thus, not having a need for U does not mean that there is no Mott-like transition to consider.

(v) As is often the case, the nature of the exchange-correlation (XC) functional in DFT can affect some results, depending on the ability of the XC to create compact orbitals. For the four false metals considered in this paper, the use of advanced XC functionals such as SCAN[27] (without U as an add-on) or is by itself not enough to secure a proper above-Neel insulating PM state) unless one permits spin-related symmetry breaking first, such as polymorphous distribution. It is essential to use an expanded (super) cell that enables symmetry lowering and to provide an initial nudge. Otherwise, even good XC functionals could fail qualitatively in predicting the correct metal vs insulator character.

## II. Methods: Examining magnetic and structural symmetry breaking by avoiding the use of virtual averaged configurations

To describe the electronic properties of the compounds, we use symmetry broken density functional theory approximation allowing the existence of structural and magnetic m-DOFs as long as their formation lowers the internal energy of the system. For magnetically ordered phases (e.g., ferromagnets (FM) and antiferromagnets (AFM)), the lowest magnetic orders identified by Varignon et al.[18,19,28] are used. For the PM phases, we utilize the spin polymorphous description - *spin* special quasirandom structure (SQS)[29] is generated for spin-up and spin-down local magnetic motifs with a global zero magnetic moment using a 160-atom supercell. Such a polymorphous description corresponds to the high-temperature limit of a random spin paramagnet [16,30]. There are a few important consequences of the application of the spin polymorphous model: (i) spin-polymorphous system has significantly lower energy compared to nonmagnetic approximation in the system; (ii) while the nonmagnetic system can often be approximated as a compound with specifically defined Wyckoff positions, the spin-polymorphous model has local structural symmetry making each site structurally and magnetically unique; (iii) each transition metal atom has magnetic moments. The first-principles calculations are carried out using plane-wave DFT as implemented in the Vienna Ab Initio Simulation Package (VASP)[31-33]. SCAN[27] meta GGA functional (no +U correction) or PBEsol[34] with U correction applied on d-electrons of transition metals as introduced by Dudarev et al.[35] are used to describe exchange-correlation interaction. The cutoff energies for the plane-wave basis are set to 500 eV for final calculations and 550 eV for volume relaxation. Atomic relaxations are performed until the internal forces are smaller than 0.01 eV/Å unless specified. Analysis of structural properties and visualization of computed results are performed using Vesta[36] and pymatgen library[37].

## III. Results

### A. $d^1$ perovskite: what breaks the degeneracy creating a true insulator (LaTiO$_3$) and what retains the intrinsic metallic state (SrVO$_3$)

LaTiO$_3$ is an orthorhombic (space group: Pnma, tolerance factor: 0.93) insulator both below the Néel temperature (T$_N$=146 K[38]), where it exists as the AFM α phase, and above the Néel temperature, where it exists as the PM β phase[39] in the same crystal structure (Fig. 1a). With respect to reference high symmetry Pm-3m structure (not observed experimentally for LaTiO$_3$), the ground state structure has distinct symmetry lowering inclining the TiO$_6$ octahedron around a [110]$_c$ axis, and subsequently executing a rotation about the c axis[38]. In the strongly correlated literature[9,10], the observed insulating character of the PM β phase was argued to result from strong electron correlation-absent in band theory that incorrectly predicted this phase to be a "false metal". In contrast to such a purely electronic degree of freedom explanation, mean-field symmetry broken polymorphous electronic structure works emphasize the effect of m-DOF. That the α phase is insulating has been explained by the existence of AFM long-range order in the presence of energy-lowering octahedra tilting/distortion[28]. However, the fact that the PM β phase is also an insulator was initially unexpected as this phase lacks the (spin) LRO that enabled gapping of the α phase. The transition between a low-temperature spin-ordered α phase to a PM β does not necessarily involve a sudden disappearance of all local moments. Indeed, finite local moments are observed in PM phases, as evidenced by experimental measurements of magnetic pair distribution function[40,41] or, more generally, unexpected magnetic responses of PM phases.[42-49]

***Insulating LaTiO$_3$ ($t^1e^0$): paramagnet within symmetry broken polymorphous electronic structure:*** For PM β LaTiO$_3$ (same also stays for PM β YTiO$_3$), the insulating nature of the compound can be well captured by accounting for the energy lowering structural and magnetic symmetry breaking. For instance, our calculations demonstrate that the spin polymorphous model has internal energy of 61 meV/atom lower than that for nonmagnetic phase. Figures 1b and 1c show the density of states (DOS) of the β-PM phase in the same supercell with and without spin symmetry breaking. The results show that both structural and magnetic symmetry breaking is needed to describe the electronic structure of the system (i.e., magnetic or structural symmetry breaking alone does not open the gap, Fig. 1a). In the PM β phase, the local structural and magnetic patterns are comparable to those in the AFM α phase. However, the key difference lies in the absence of spin magnetic moment ordering. This demonstrates that while the PM β phase shares similarities with the AFM α phase, it does not exhibit long-range spin order (as it will be seen below, this is the common tendency for range of the compound discussed in this work). Moreover, only spin symmetry breaking (i.e., ignoring the structural motifs as compared to ideal high-symmetry Pm-3m perovskite structure) or only structural symmetry breaking without magnetic symmetry breaking (i.e., using naïve nonmagnetic approximation of the paramagnet) is not sufficient for band gap opening within DFT. These results question if DMFT predictions based on the Hamiltonian mapping of primitive cells can sufficiently well capture the main physical phenomena originating from the local symmetry breaking without accounting for the existence of distribution of m-DOFs.

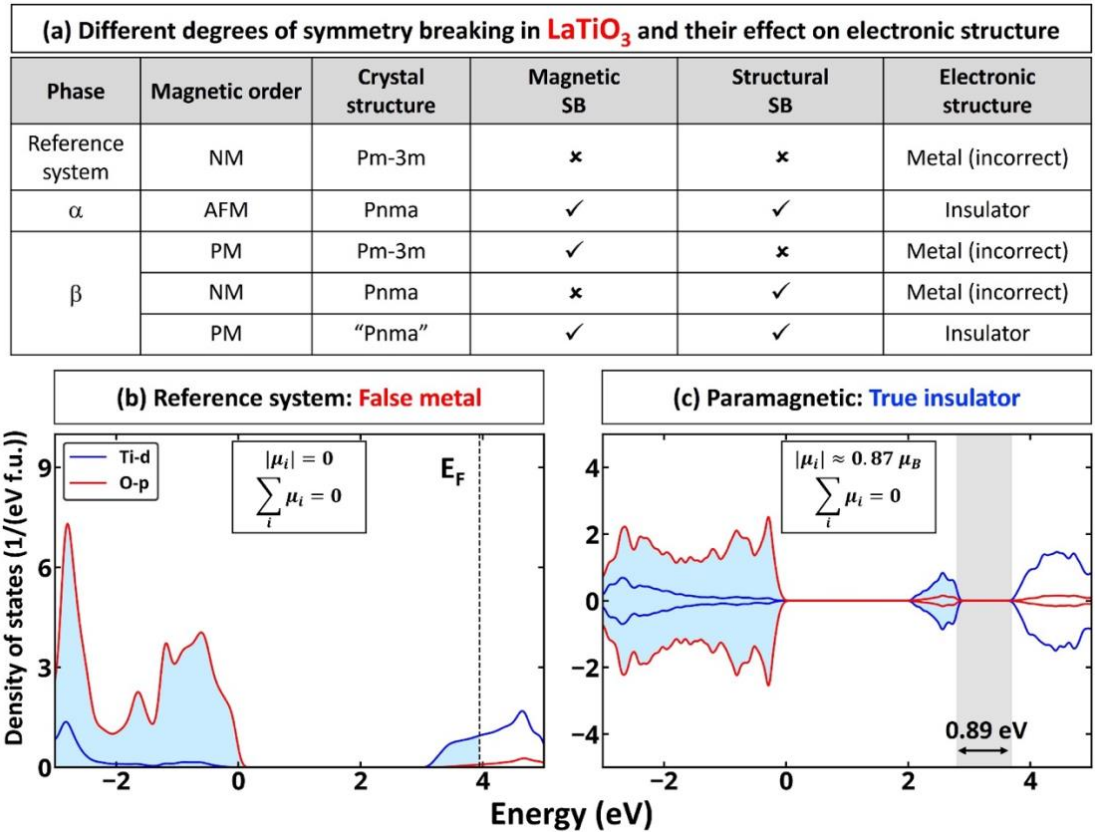

**Figure 1:** (a) Summary of different degrees of symmetry breaking (SB) and their effect on the electronic structure in LaTiO$_3$. The "reference system" both below or above the Néel transition is defined here as a monomorphous cubic structure without any local motifs that could break symmetry, hence being a (possibly false) metal. Electronic structure of β-PM LaTiO$_3$ as described with (b) nonmagnetic and (c) spin polymorphous model where both structural and spin symmetry breaking is allowed. The results are presented for PBEsol+U with a U value of 2.5 eV. The inset show details on local and total magnetic moments in the system. The space group in quotation marks references to a global average structure. Occupied states are shown as shadowed.

***Persistent metal SrVO$_3$ d$^1$(t$^1$e$^0$): paramagnet within symmetry broken electronic structure:*** This system was treated previously by the same methods as described here[19,24] and is included for completeness. The importance of both structural and spin symmetry breaking on electronic properties of PM β LaTiO$_3$ implies that if the structural symmetry breaking or magnetic symmetry breaking is suppressed because of an intrinsic tendency of the system or more generally the effect of external knobs (e.g., temperature, pressure) on the material, the (d,d*) paramagnets can be metallic. For instance, SrVO$_3$ (space group: Pm-3m) and CaVO$_3$ (space group: Pnma) are PM degenerate gapped metals with the Fermi level in the principal conduction band until the lowest temperature tried.[6,7] Both of these compounds can open the band gap due to the critical structural symmetry breaking within the spin polymorphous model[19], however, the structural symmetry breaking needed to open the gap are energetically unfavorable without accounting for external knobs. The fact is that both SrVO$_3$ and CaVO$_3$ have a tolerance factor of close to one (1.01 and 0.98, respectively), which does not allow the compounds to develop sufficiently strong symmetry breaking. We note, however, that using the spin polymorphous model to describe the electronic properties of such compounds is still crucial, as accounting for the distribution of local magnetic motifs is needed to describe the electronic properties of the compounds. For instance, Wang et al. demonstrated that the mass enhancement in SrVO$_3$ can be explained by accounting for spin symmetry breaking without including for any dynamic correlation effect.[24]

## B. LaVO$_3$: the d$^2$ (t$^2$e$^0$): electron perovskite that breaks the symmetry

This d$^2$ system was treated before (i.e., Refs. [18,19,28]) and here we summarize the salient results to place the other d$^1$, d$^3$, and d$^4$ perovskites into perspective. The ground state structure of LaVO$_3$ (α phase, tolerance factor: 0.95) is the AFM monoclinic P2$_1$/b structure that turns into the PM orthorhombic Pnma structure (β phase) around 140 K.[50] Both phases are insulators with confirmed Q$^{2+}$ Jahn-Teller (JT) distortion.[28] While a metallic behavior is expected for LaVO$_3$ (t$_{2g}^2$e$_g^0$) with its two electrons distributed in three t$_{2g}$ partners, they are experimentally found to be insulators.[51] For the ground state structure, the DFT calculations with SCAN or PBEsol+U functional confirmed that the AFM-C magnetic configuration with monoclinic (P2$_1$/b) structure is the lowest energy structure with band gap energy of 0.42 eV using PBEsol+U, U=3.5 eV[28] or 0.78 eV using SCAN[18]. We note, however, that using soft-XC functionals is unable to predict the true insulating nature.[5] The PM orthorhombic structure modeled by a monomorphous nonmagnetic spin configuration based on the averaged spin structure is understandably found to be a "False metal". This does not reflect on a shortcoming of the DFT but rather on a misrepresentation of PM phases. Indeed the NM false metal configuration has enormously higher energy (by 1167 meV/f.u.) than the spin polymorphous of the same crystallographic structure.[19] What makes LaVO$_3$ different from some other oxides is that rotations plus antipolar displacements of ions with respect to ideal Wyckoff positions in the high-symmetry cubic structure are sufficient to produce an insulating state since the $Q^{2-}$ JT mode is not important for the gap opening. We note, however, that similar to the β phase of LaTiO$_3$ band gap opening in both α and β phases of LaVO$_3$ requires an accounting of both structural and spin symmetry breaking, i.e., accounting of only one does not result in band gap opening.

**C.      SrMnO$_3$: the d$^3$ (t$^2$e$^1$) electron perovskite that breaks symmetry to become an insulator**

SrMnO$_3$ is an example of (d, d*) insulator that at low temperature exists in an antiferromagnetically ordered cubic(Pm-3m, tolerance factor: 1.03) structure (α phase) and exhibits the Néel transition to β phase at the temperature of about 260 K to PM cubic insulator.[52-54] We note however that depending on synthesis conditions, orthorhombic (C222$_1$) AFM, PM β orthorhombic (C222$_1$), and PM β tetragonal (P6$_3$/mmc) can be observed.[55] Recently Gull et al.[12] suggested that none of the conventional theoretical approaches could obtain an *insulating* result for PM cubic SrMnO$_3$. Instead, exceedingly complex methods  (with the self-energy embedding theory method[56-58] with using GW[59] as the weakly correlated "outer" method for all orbitals and exact diagonalization as the "inner" quantum impurity solver for the correlated orbitals) are *required.*

***True insulating state in α and β phases of SrMnO$_3$:*** Using the mean field like DFT but without ignoring symmetry breaking in describing the cubic PM phase, we observed experimentally expected cubic PM insulating phase without accounting for any dynamic correlation. Specifically, we confirm that viewing a PM phase as one that not only has a global zero magnetization but also has each atom have zero local moments (a monomorphous nonmagnetic approximation) will necessarily lead in a cubic phase to a false metallic state (Fig. 2b). Replacing this rather naïve monomorphous description with a polymorphous description still has the macroscopic cubic shape but each Mn atom is not constrained to be nonmagnetic –naturally lowers the total energy and gives an insulating phase. The calculated band gap is 1.12 eV using the SCAN exchange-correlation functional. The energy of the spin polymorphous system is 460 meV/atom lower than that for NM symmetry unbroken description. In contrast to monomorphous nonmagnetic approximation, within the spin polymorphous approach, we find that there is a distribution of local magnetic moments on Mn atom with the average absolute value of magnetic moment on each side of 2.63±0.04 $\mu_B$. While the structural symmetry breaking provided by local distortions of the individual octahedra is energy lowering, this is not the reason for the band gap opening. Without magnetic symmetry breaking, the structural symmetry breaking in cubic SrMnO$_3$ cannot open the gap. The band gap opening originates from the magnetic symmetry breaking, which is already by itself sufficient to open the band gap. We note however that structural symmetry breaking is still an important factor as it can affect the absolute value of band gap energy. These results are similar to those found for cubic PM CaMnO$_3$[19], where the authors demonstrated that band gap opening is due to crystal-field splitting and demonstrate that crystal-field splitting dependent strongly on the degree of magnetic symmetry breaking.

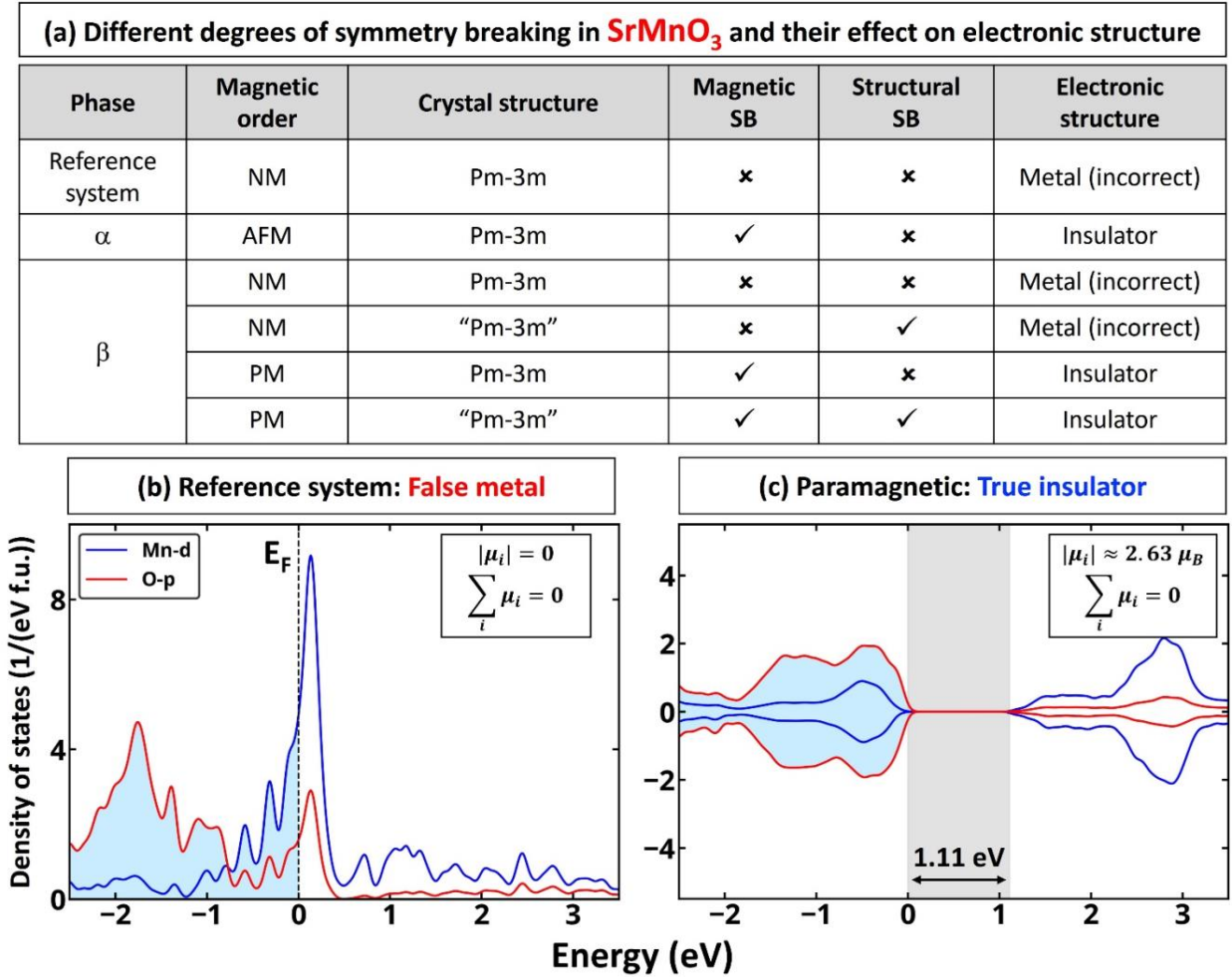

**Figure 2.** (a) Summary of different degrees of symmetry breaking (SB) and their effect on the electronic structure in SrMnO$_3$. The "reference system" both below or above the Néel transition is defined here as a monomorphous cubic structure without any local motifs that could break symmetry. Hence, in this reference state the system is a metal. Electronic structure of β-PM SrMnO$_3$ as described with (b) nonmagnetic and (c) spin polymorphous model where both structural and spin symmetry breaking is allowed. The results are presented for SCAN. The inset show details on local and total magnetic moments in the system. The space group in quotation marks references to the global average structure. Occupied states are shown as shadowed.

**D.　LaMnO$_3$: the d$^4$ (t$^3$e$^1$) electron perovskite that breaks the symmetry**

The ground state LaMnO$_3$ is an AFM insulator with an orthorhombic structure (α phase, Pnma, tolerance factor: 0.94), in which structural symmetry breaking is represented by octahedral tilting occurs and distinct octahedra distortion. Recently, such structural changes have been characterized as pseudo Jahn-Teller distortion[28]. Upon heating, the AFM insulator turns in the PM insulator around 140 K (Néel temperature) without structural phase transition (β phase, space group: Pnma).[60] At a temperature above 780 K, the β phase turns into γ phase – cubic (Pm-3m) PM polaron conductors.[60,61] We note that despite some crystallographic data confirming the phase transition, there are clear indication of serving local structural symmetry breaking even above the phase transition temperature in the $\gamma$ phase[62].

***True insulating state in $\alpha$ and $\beta$ phases*** **and** ***true metallic state of $\gamma$ phase of LaMnO₃***: For the orthorhombic $\alpha$ phase of LaMnO₃, by using AFM configuration, we indeed find the insulating state with SCAN band gap 0.49 eV (see Fig. 3(a)). These results are consistent with previous experimental and theoretical results [18,19,28]. Moreover, the calculated local spin moment on Mn atom is 3.66 $\mu_B$, close to the value measured in the experiment (3.7 $\mu_B$) [63]. We note that the description of the insulating state in the $\alpha$ phase requires accounting for structural distortion present in the orthorhombic phase. If the $\alpha$ phase is approximated as an ideal high symmetry Pm-3m structure (i.e., one often used in the correlation models), the magnetic symmetry breaking (i.e., AFM) is unable to open the gap. These results thus clearly highlight the importance of structural symmetry breaking (specifically pseudo-Jahn-Teller distortion) in the band gap opening.

While the origin of band gap opening in $\alpha$ phase of LaMnO₃ is rather well understood, historically, the $\beta$ phase attracted the most interest in the literature on correlated materials[13] using global average nonmagnetic configuration (zero spin moment on each Mn atom) is found to be a "false metal". Hence, the inability to describe the insulating state of the $\beta$ phase of LaMnO₃ within naïve DFT approximation historically has been used as motivation to develop post-DFT methods (like DMFT). For the β phase, the spin-polymorphous model can describe the insulating state when structural and magnetic symmetry breaking are accounted for (Fig. 3a,b). The spin polymorphous system has the distribution of local magnetic moments in the Mn site with a magnetic moment of 3.66±0.01 $\mu_B$ and substantially lower internal energy (as compared to the naïve nonmagnetic approximation of the paramagnet). These results thus demonstrate that the gapping of the $\beta$ phase can be described by the accounting of the formation of m-DOFs that lower internal energy without accounting for any dynamic correlation. Importantly, in contrast to cubic PM SrMnO₃, accounting for only one m-DOF (i.e., structural symmetry breaking or magnetic symmetry breaking) is insufficient to open the gap (Fig. 3a).

What make the case of LaMnO₃ interesting is the fact that $\gamma$ phase is metallic with spin and structure symmetry breaking accounted for (see Fig. 3a), which is in good agreement with experimentally observed polaron conductor behavior [60,61]. Since the Mn magnetic moments for the $\gamma$ phase and the $\beta$ phase are close (3.63 $\mu_B$ for the former and 3.66 $\mu_B$ for the later), we conclude that insulating states (similar to case of PM SrVO₃ and CaVO₃) need sufficient structure symmetry breaking seen in $\beta$ phase which is however not developed in $\gamma$ phase of LaMnO₃ as the result of energy lowering formation of m-DOFs.

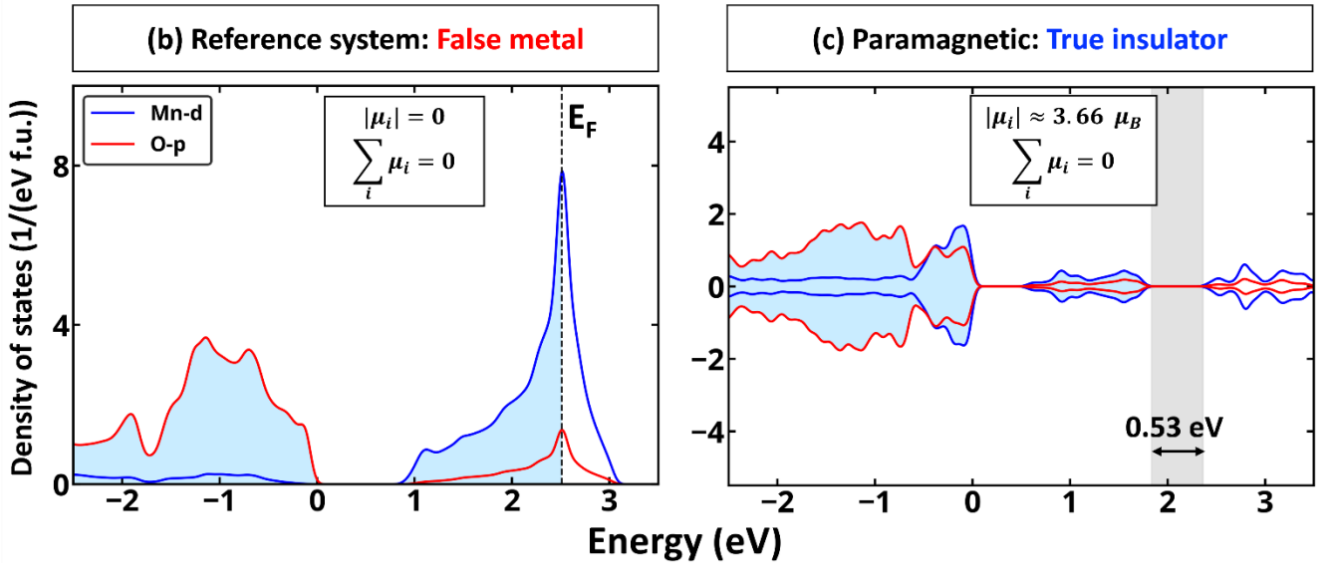

**Figure 3.** Summary of different degrees of symmetry breaking (SB) and their effect on the electronic structure in LaMnO$_3$. The "reference system" both below or above the Néel transition is defined here as a monomorphous cubic structure without any local motifs that could break symmetry. Electronic structure of β-PM LaMnO$_3$ as described with (b) nonmagnetic and (c) spin polymorphous model where both structural and spin symmetry breaking is allowed. The results are presented for SCAN. The inset show details on local and total magnetic moments in the system. The space group in quotation marks references to the global average structure. Occupied states are shown as shadowed.

**IV. Summary and Discussion:**

We have demonstrated that the electronic properties of LaTiO$_3$, LaVO$_3$, SrMnO$_3$, and LaMnO$_3$ compounds, previously referred to as "false metals", can be described using DFT both below and above the Néel temperature. Specifically, such accurate description requires consideration for three key factors:

(i) one should not average use the average unit cell. Instead, we should expand it to a supercell with spin and structural SQS, which permits more degrees of freedom. For instance, for materials like LaTiO$_3$ and LaVO$_3$, acknowledging this spin and structural symmetry disruption lets us articulate their insulating characteristics. Similarly, the insulating state in the beta phase of LaMnO$_3$ can be characterized when factoring in the spin and structural symmetry breaking.

(ii) we should consider advanced XC functional usage. While this study does not prioritize establishing the superiority of SCAN or PBEsol+U in calculating the electronic properties of the discussed compounds, it does demonstrate that they, along with others like hybrid [64-67] or r$^2$SCAN[68] functionals, can yield accurate electronic structures if they allow for energy-lowering symmetry disruptions.

(iii) there should be an allowance for collaboration between different modes of microscopic degrees of freedom. For instance, in SrMnO$_3$, the band gap opening stems from the magnetic symmetry disruption, while the structural symmetry disruption modifies the overall band gap energy value.

This three-pronged approach not only allows us to reproduce the Néel temperature for paramagnetic materials[30] accurately and describe the metal-insulator transition[23] when the weakening of structural symmetry disruption due to temperature is taken into account. Furthermore, it also provides a thorough understanding of the electronic properties of these materials without the necessity for more complex, post-DFT methods. The data set of materials discussed here is limited, but it emphasizes that a polymorphous network and its influence on the electronic properties of quantum materials is a broad topic. Therefore, it is critical to avoid making premature conclusions about strong electron-electron correlation in pseudo-metal syndrome cases. Instead, the three-pronged strategy we propose – namely expanding to a supercell, utilizing advanced XC functionals, and allowing collaborations between different modes of microscopic degrees of freedom – can satisfactorily explain the lack of false metals in numerous cases.

*Acknowledgment:* The work on magnetic symmetry breaking and electronic structure calculations was supported by U.S. Department of Energy, Office of Science, Basic Energy Sciences, Materials Sciences and Engineering Division within grant DE-SC0010467 to CU Boulder, while using resources of the National Energy Research Scientific Computing Center, which is supported by the Office of Science of the U.S. Department of Energy. The authors acknowledge also use of computational resources located at the National Renewable Energy Laboratory and sponsored by the Department of Energy's Office of Energy Efficiency and Renewable Energy. Work on structural and spin symmetry breaking was supported with the support of NSF-DMREF, grant number 1921949. The authors acknowledge using Extreme Science and Engineering Discovery Environment (XSEDE) supercomputer resources, supported by the National Science Foundation, grant number ACI-1548562.

**References:**
[1] J. H. d. Boer and E. J. W. Verwey, Semi-conductors with partially and with completely filled 3d-lattice bands, Proceedings of the Physical Society, **49**, 59 (1937).
[2] N. F. Mott, The basis of the electron theory of metals, with special reference to the transition metals, Proceedings of the Physical Society. Section A, **62**, 416 (1949).

[3] N. F. Mott and R. Peierls, Discussion of the paper by de Boer and Verwey, Proceedings of the Physical Society, **49**, 72 (1937).
[4] J. Hubbard, Electron correlations in narrow energy bands, Proc. R. Soc. London, Ser. A, **276**, 238 (1963).
[5] O. I. Malyi and A. Zunger, False metals, real Insulators, and degenerate gapped metals, Appl. Phys. Rev., **7**, 041310 (2020).
[6] L. Zhang, Y. J. Zhou, L. Guo, W. W. Zhao, A. Barnes, H. T. Zhang, C. Eaton, Y. X. Zheng, M. Brahlek, H. F. Haneef, N. J. Podraza, M. H. W. Chan, V. Gopalan, K. M. Rabe, and R. Engel-Herbert, Correlated metals as transparent conductors, Nat. Mater., **15**, 204 (2016).
[7] M. Takizawa, M. Minohara, H. Kumigashira, D. Toyota, M. Oshima, H. Wadati, T. Yoshida, A. Fujimori, M. Lippmaa, M. Kawasaki, H. Koinuma, G. Sordi, and M. Rozenberg, Coherent and incoherent d band dispersions in $SrVO_3$, Phys. Rev. B, **80**, 235104 (2009).
[8] J. A. Alonso, J. L. García-Muñoz, M. T. Fernández-Díaz, M. A. G. Aranda, M. J. Martínez-Lope, and M. T. Casais, Charge disproportionation in $RNiO_3$ perovskites: simultaneous metal-insulator and structural transition in $YNiO_3$, Physical Review Letters, **82**, 3871 (1999).
[9] E. Pavarini, S. Biermann, A. Poteryaev, A. I. Lichtenstein, A. Georges, and O. K. Andersen, Mott transition and suppression of orbital fluctuations in orthorhombic $3d^1$ perovskites, Phys. Rev. Lett., **92**, 176403 (2004).
[10] E. Pavarini, A. Yamasaki, J. Nuss, and O. K. Andersen, How chemistry controls electron localization in $3d^1$ perovskites: a Wannier-function study, New J. Phys., **7**, 188 (2005).
[11] M. De Raychaudhury, E. Pavarini, and O. K. Andersen, Orbital fluctuations in the different phases of $LaVO_3$ and $YVO_3$, Phys. Rev. Lett., **99**, 126402 (2007).
[12] C.-N. Yeh, S. Iskakov, D. Zgid, and E. Gull, Electron correlations in the cubic paramagnetic perovskite $Sr(V,Mn)O_3$: Results from fully self-consistent self-energy embedding calculations, Phys. Rev. B, **103**, 195149 (2021).
[13] E. Pavarini, Solving the strong-correlation problem in materials, La Rivista del Nuovo Cimento, **44**, 597 (2021).
[14] A. Zunger, Bridging the gap between density functional theory and quantum materials, Nature Computational Science, **2**, 529 (2022).
[15] J. Varignon, O. I. Malyi, and A. Zunger, Dependence of band gaps in d-electron perovskite oxides on magnetism, Phys. Rev. B, **105**, 165111 (2022).
[16] D. Gambino, O. I. Malyi, Z. Wang, B. Alling, and A. Zunger, Density functional description of spin, lattice, and spin-lattice dynamics in antiferromagnetic and paramagnetic phases at finite temperatures, Physical Review B, **106**, 134406 (2022).
[17] O. I. Malyi, X.-G. Zhao, A. Bussmann-Holder, and A. Zunger, Local positional and spin symmetry breaking as a source of magnetism and insulation in paramagnetic $EuTiO_3$, Phys. Rev. Mater., **6**, 034604 (2022).
[18] J. Varignon, M. Bibes, and A. Zunger, Mott gapping in 3d $ABO_3$ perovskites without Mott-Hubbard interelectronic repulsion energy U, Phys. Rev. B, **100**, 035119 (2019).
[19] J. Varignon, M. Bibes, and A. Zunger, Origin of band gaps in 3d perovskite oxides, Nat. Commun., **10**, 1658 (2019).
[20] G. Trimarchi, Z. Wang, and A. Zunger, Polymorphous band structure model of gapping in the antiferromagnetic and paramagnetic phases of the Mott insulators MnO, FeO, CoO, and NiO, Phys. Rev. B, **97**, 035107 (2018).

[21]   X. Xu, C. Randorn, P. Efstathiou, and J. T. S. Irvine, A red metallic oxide photocatalyst, Nat. Mater., **11**, 595 (2012).

[22]   M. T. Casais, J. A. Alonso, I. Rasines, and M. A. Hidalgo, Preparation, neutron structural study and characterization of $BaNbO_3$ - a Pauli-like metallic perovskite, Mater. Res. Bull., **30**, 201 (1995).

[23]   O. I. Malyi and A. Zunger, Rise and fall of Mott insulating gaps in $YNiO_3$ paramagnets as a reflection of symmetry breaking and remaking, Physical Review Materials, **7**, 044409 (2023).

[24]   Z. Wang, O. I. Malyi, X. Zhao, and A. Zunger, Mass Enhancement in 3d and s-p Perovskites from Symmetry Breaking, Phys. Rev. B, **103** 165110 (2021).

[25]   X.-G. Zhao, Z. Wang, O. I. Malyi, and A. Zunger, Effect of static local distortions vs. dynamic motions on the stability and band gaps of cubic oxide and halide perovskites, Materials Today, **49**, 107 (2021).

[26]   X.-G. Zhao, G. M. Dalpian, Z. Wang, and A. Zunger, Polymorphous nature of cubic halide perovskites, Physical Review B, **101**, 155137 (2020).

[27]   J. Sun, A. Ruzsinszky, and J. P. Perdew, Strongly constrained and appropriately normed semilocal density functional, Phys. Rev. Lett., **115**, 036402 (2015).

[28]   J. Varignon, M. Bibes, and A. Zunger, Origins versus fingerprints of the Jahn-Teller effect in d-electron $ABX_3$ perovskites, Phys. Rev. Res., **1**, 033131 (2019).

[29]   A. Zunger, S. H. Wei, L. G. Ferreira, and J. E. Bernard, Special quasirandom structures, Phys. Rev. Lett., **65**, 353 (1990).

[30]   J. L. Du, O. I. Malyi, S. L. Shang, Y. Wang, X. G. Zhao, F. Liu, A. Zunger, and Z. K. Liu, Density functional thermodynamic description of spin, phonon and displacement degrees of freedom in antiferromagnetic-to-paramagnetic phase transition in $YNiO_3$, Materials Today Physics, **27**, 100805 (2022).

[31]   G. Kresse and J. Hafner, Ab initio molecular dynamics for liquid metals, Physical Review B, **47**, 558 (1993).

[32]   G. Kresse and J. Furthmüller, Efficiency of ab-initio total energy calculations for metals and semiconductors using a plane-wave basis set, Computational Materials Science, **6**, 15 (1996).

[33]   G. Kresse and J. Furthmüller, Efficient iterative schemes for ab initio total-energy calculations using a plane-wave basis set, Phys. Rev. B, **54**, 11169 (1996).

[34]   J. P. Perdew, A. Ruzsinszky, G. I. Csonka, O. A. Vydrov, G. E. Scuseria, L. A. Constantin, X. Zhou, and K. Burke, Restoring the density-gradient expansion for exchange in solids and surfaces, Phys. Rev. Lett., **100**, 136406 (2008).

[35]   S. L. Dudarev, G. A. Botton, S. Y. Savrasov, C. J. Humphreys, and A. P. Sutton, Electron-energy-loss spectra and the structural stability of nickel oxide: An LSDA+U study, Phys. Rev. B, **57**, 1505 (1998).

[36]   K. Momma and F. Izumi, VESTA 3 for Three-dimensional visualization of crystal, volumetric and morphology data, J. Appl. Crystallogr., **44**, 1272 (2011).

[37]   S. P. Ong, W. D. Richards, A. Jain, G. Hautier, M. Kocher, S. Cholia, D. Gunter, V. L. Chevrier, K. A. Persson, and G. Ceder, Python Materials Genomics (Pymatgen): A robust, open-source python library for materials analysis, Comp. Mater. Sci., **68**, 314 (2013).

[38]   M. Cwik, T. Lorenz, J. Baier, R. Müller, G. André, F. Bourée, F. Lichtenberg, A. Freimuth, R. Schmitz, E. Müller-Hartmann, and M. Braden, Crystal and magnetic structure of $LaTiO_3$: Evidence for nondegenerate $t_{2g}$ orbitals, Phys. Rev. B, **68**, 060401 (2003).

[39] J. Hemberger, H. A. K. von Nidda, V. Fritsch, J. Deisenhofer, S. Lobina, T. Rudolf, P. Lunkenheimer, F. Lichtenberg, A. Loidl, D. Bruns, and B. Büchner, Evidence for Jahn-Teller distortions at the antiferromagnetic transition in LaTO$_3$, Phys. Rev. Lett., **91**, 066403 (2003).

[40] B. Frandsen, X. Yang, and S. J. Billinge, Magnetic pair distribution function analysis of local magnetic correlations, Acta Crystallographica Section A: Foundations and Advances, **70**, 3 (2014).

[41] B. A. Frandsen, M. Brunelli, K. Page, Y. J. Uemura, J. B. Staunton, and S. J. L. Billinge, Verification of Anderson superexchange in MnO via magnetic pair distribution function analysis and ab initio theory, Phys. Rev. Lett., **116**, 197204 (2016).

[42] Z. Guguchia, H. Keller, R. K. Kremer, J. Köhler, H. Luetkens, T. Goko, A. Amato, and A. Bussmann-Holder, Spin-lattice coupling induced weak dynamical magnetism in EuTiO$_3$ at high temperatures, Phys. Rev. B, **90**, 064413 (2014).

[43] Z. Guguchia, H. Keller, J. Köhler, and A. Bussmann-Holder, Magnetic field enhanced structural instability in EuTiO$_3$, J. Phys.: Condens. Matter, **24**, 492201 (2012).

[44] A. Bussmann-Holder, J. Köhler, K. Roleder, Z. Guguchia, and H. Keller, Unexpected magnetism at high temperature and novel magneto-dielectric-elastic coupling in EuTiO$_3$ : A critical review, Thin Solid Films, **643**, 3 (2017).

[45] G. Gregori, J. Köhler, J. F. Scott, and A. Bussmann-Holder, Hidden magnetism in the paramagnetic phase of EuTiO$_3$, J. Phys.: Condens. Matter, **27**, 496003 (2015).

[46] Z. Guguchia, Z. Salman, H. Keller, K. Roleder, J. Köhler, and A. Bussmann-Holder, Complexity in the structural and magnetic properties of almost multiferroic EuTiO$_3$, Phys. Rev. B, **94**, 220406 (2016).

[47] P. Pappas, M. Calamiotou, M. Polentarutti, G. Bais, A. Bussmann-Holder, and E. Liarokapis, Magnetic field driven novel phase transitions in EuTiO$_3$, arXiv:2103.04742 [cond-mat], (2021).

[48] T. Katsufuji and H. Takagi, Coupling between magnetism and dielectric properties in quantum paraelectric EuTiO$_3$, Phys. Rev. B, **64**, 054415 (2001).

[49] A. Bussmann-Holder, Z. Guguchia, J. Köhler, H. Keller, A. Shengelaya, and A. R. Bishop, Hybrid paramagnon phonon modes at elevated temperatures in EuTiO$_3$, New J. Phys., **14**, 093013 (2012).

[50] D. J. Lovinger, M. Brahlek, P. Kissin, D. M. Kennes, A. J. Millis, R. Engel-Herbert, and R. D. Averitt, Influence of spin and orbital fluctuations on Mott-Hubbard exciton dynamics in LaVO$_3$ thin films, Phys. Rev. B, **102**, 115143 (2020).

[51] S. Miyasaka, Y. Okimoto, M. Iwama, and Y. Tokura, Spin-orbital phase diagram of perovskite-type RVO$_3$ (R=rare-earth ion or Y), Phys. Rev. B, **68**, 100406 (2003).

[52] O. Chmaissem, B. Dabrowski, S. Kolesnik, J. Mais, J. D. Jorgensen, and S. Short, Structural and magnetic phase diagrams of La$_{1-x}$Sr$_x$MnO$_3$ and Pr$_{1-y}$Sr$_y$MnO$_3$, Phys. Rev. B, **67**, 094431 (2003).

[53] R. K. Hona and F. Ramezanipour, Effect of the oxygen vacancies and structural order on the oxygen evolution activity: a case study of SrMnO$_{3−δ}$ featuring four different structure types, Inorg. Chem., **59**, 4685 (2020).

[54] T. Takeda and S. Ōhara, Magnetic structure of the cubic perovskite self-energy embedding theory (SEET) for periodic systems type SrMnO$_3$, J. Phys. Soc. Jpn., **37**, 275 (1974).

[55] A. Daoud-Aladine, C. Martin, L. C. Chapon, M. Hervieu, K. S. Knight, M. Brunelli, and P. G. Radaelli, Structural phase transition and magnetism in hexagonal SrMnO$_3$ by magnetization measurements and by electron, x-ray, and neutron diffraction studies, Phys. Rev. B, **75**, 104417 (2007).

[56] A. A. Kananenka, E. Gull, and D. Zgid, Systematically improvable multiscale solver for correlated electron systems, Phys. Rev. B, **91**, 121111 (2015).

[57]   D. Zgid and E. Gull, Finite temperature quantum embedding theories for correlated systems, New J. Phys., **19**, 023047 (2017).

[58]   A. A. Rusakov, S. Iskakov, L. N. Tran, and D. Zgid, Self-energy embedding theory (SEET) for periodic systems, J. Chem. Theory Comput., **15**, 229 (2019).

[59]   F. Aryasetiawan and O. Gunnarsson, The GW method, Rep. Prog. Phys., **61**, 237 (1998).

[60]   Y. Murakami, J. P. Hill, D. Gibbs, M. Blume, I. Koyama, M. Tanaka, H. Kawata, T. Arima, Y. Tokura, K. Hirota, and Y. Endoh, Resonant X-ray scattering from orbital ordering in $LaMnO_3$, Phys. Rev. Lett., **81**, 582 (1998).

[61]   J. S. Zhou and J. B. Goodenough, Paramagnetic phase in single-crystal $LaMnO_3$, Phys. Rev. B, **60**, R15002 (1999).

[62]   X. Qiu, T. Proffen, J. F. Mitchell, and S. J. L. Billinge, Orbital correlations in the pseudocubic O and rhombohedral R phases of $LaMnO_3$, Physical Review Letters, **94**, 177203 (2005).

[63]   J. B. A. A. Elemans, B. Van Laar, K. R. Van Der Veen, and B. O. Loopstra, The crystallographic and magnetic structures of $La_{1-x}Ba_xMn_{1-x}Me_xO_3$ (Me = Mn or Ti), J. Solid State Chem., **3**, 238 (1971).

[64]   Y. Jiao, E. Schröder, and P. Hyldgaard, Extent of Fock-exchange mixing for a hybrid van der Waals density functional?, The Journal of Chemical Physics, **148** (2018).

[65]   V. Shukla, Y. Jiao, J.-H. Lee, E. Schröder, J. B. Neaton, and P. Hyldgaard, Accurate nonempirical range-separated hybrid van der Waals density functional for complex molecular problems, solids, and surfaces, Physical Review X, **12**, 041003 (2022).

[66]   J. Heyd, G. E. Scuseria, and M. Ernzerhof, Hybrid functionals based on a screened Coulomb potential, The Journal of Chemical Physics, **118**, 8207 (2003).

[67]   A. V. Krukau, O. A. Vydrov, A. F. Izmaylov, and G. E. Scuseria, Influence of the exchange screening parameter on the performance of screened hybrid functionals, The Journal of Chemical Physics, **125** (2006).

[68]   J. W. Furness, A. D. Kaplan, J. Ning, J. P. Perdew, and J. Sun, Accurate and numerically efficient $r^2SCAN$ meta-generalized gradient approximation, The Journal of Physical Chemistry Letters, **11**, 8208 (2020).